\documentclass[a4paper]{article}

\usepackage{INTERSPEECH_v2}
\usepackage{graphicx}
\usepackage{amssymb,amsmath,bm}
\usepackage{textcomp}
\usepackage[utf8x]{inputenc} 

\def\vec#1{\ensuremath{\bm{{#1}}}}

\ninept

\title{Gate Activation Signal Analysis for Gated Recurrent Neural Networks \\ and Its Correlation with Phoneme Boundaries}

\makeatletter
\def\name#1{\gdef\@name{#1\\}}
\makeatother \name{\em Yu-Hsuan Wang, Cheng-Tao Chung, Hung-yi Lee}

\address{College of Electrical Engineering and Computer Science, National Taiwan University}
\email {\small \tt r04922167@ntu.edu.tw, f01921031@ntu.edu.tw, hungyilee@ntu.edu.tw}
\begin{document}

  \maketitle

    \begin{abstract}
In this paper we analyze the gate activation signals inside the gated recurrent neural networks, and find the temporal structure of such signals is highly correlated with the phoneme boundaries. This correlation is further verified by a set of experiments for phoneme segmentation, in which better results compared to standard approaches were obtained.
    \newline
    \end{abstract}

    
    \section{Introduction}
        
    Deep learning has achieved great success in many areas~\cite{hinton2012deep}\cite{hermansky2000tandem}\cite{network1989handwritten}. For problems related to sequential data such as audio, video and text, significant improvements have been achieved by Gated Recurrent Neural Networks (GRNN), in which the hidden neurons form a directed cycle suitable for processing sequential data~\cite{DBLP:journals/corr/YeungRMF15}\cite{DBLP:journals/corr/LampleBSKD16}\cite{chung2016audio}\cite{adi2016sequence}. In addition to taking the neural network outputs to be used in the target applications, internal signals within the neural networks were also found useful. A good example is the bottleneck features~\cite{grezl2007probabilistic}\cite{yu2011improved}. 
    
    On the other hand, in the era of big data, huge quantities of unlabeled speech data are available but difficult to annotate, and unsupervised approaches to effectively extract useful information out of such unlabeled data are highly attractive~\cite{lee2012nonparametric}\cite{chung2013unsupervised}. Autoencoder structures have been used for extracting bottleneck features~\cite{gehring2013extracting}, while GRNN with various structures can be learned very well without labeled data. As one example, the outputs of GRNN learned in an unsupervised fashion have been shown to carry phoneme boundary information and used in phoneme segmentation~\cite{drexler2016deep}\cite{DBLP:journals/corr/MichelRTD16}.
    
    In this paper, we try to analyze the gate activation signals (GAS) in GRNN, which are internal signals within such networks. We found such signals have temporal structures highly related to the phoneme boundaries, which was further verified by phoneme segmentation experiments.

    \section{Approaches}
     
    \subsection{Gate Activation Signals (GAS) for LSTM and GRU}
        Recurrent neural networks (RNN) are neural networks whose hidden neurons form a directed cycle. Given a sequence \textbf{x} = ($x_1$, $x_2$, ..., $x_T$), RNN updates its hidden state $\mathbf{h}_t$ at time index $t$ according to the current input $x_t$ and the previous hidden state $\mathbf{h}_{t-1}$. Gated recurrent neural networks (GRNN) achieved better performance than RNN by introducing \textbf{gates} in the units to control the information flow. Two popularly used gated units are LSTM and GRU~\cite{hochreiter1997long}\cite{cho2014properties}.

The signals within an LSTM recurrent unit are formulated as:        
         \begin{equation}
         f_{t} = \sigma(\mathbf{W}_{f}x_t + \mathbf{U}_{f}\mathbf{h}_{t-1} + \mathbf{b}_{f})
         \label{eq:f}
         \end{equation}
         \begin{equation}
         i_{t} = \sigma(\mathbf{W}_{i}x_t + \mathbf{U}_{i}\mathbf{h}_{t-1} + \mathbf{b}_{i})
         \label{eq:i}
         \end{equation}
          \begin{equation}
         \widetilde{c_{t}} = tanh(\mathbf{W}_{c}x_t + \mathbf{U}_{c}\mathbf{h}_{t-1} + \mathbf{b}_{c})
         \end{equation}
         \begin{equation}
         c_{t} = f_{t}c_{t-1} + i_{t}\widetilde{c_{t}}
         \end{equation}
         \begin{equation}
         o_{t} = \sigma(\mathbf{W}_{o}x_t + \mathbf{U}_{o}\mathbf{h}_{t-1} + \mathbf{b}_{o})
         \label{eq:o}
         \end{equation}
         \begin{equation}
         h_{t} = o_{t}tanh(c_{t})
         \end{equation}
         where $f_{t}$, $i_{t}$, $o_{t}$, $c_{t}$, $\widetilde{c_{t}}$ and $h_{t}$ are the signals over the forget gate, input gate, output gate, cell state, candidate cell state and hidden state at time $t$, respectively; $\sigma(\cdot)$ and $tanh(\cdot)$ are the sigmoid and hyperbolic tangent activation functions respectively; $\mathbf{W}_{\star}$ and $\mathbf{U}_{\star}$ are the weight matrices and $\mathbf{b}_{\star}$ are the biases.
     
        The GRU modulates the information flow inside the unit without a memory cell,
        
         \begin{equation}
         h_{t} = (1 - z_{t})h_{t-1} + z_{t}\widetilde{h}_{t}
         \end{equation} 
         \begin{equation}
         z_{t} = \sigma(\mathbf{W}_{z}x_{t} + \mathbf{U}_{z}\mathbf{h}_{t-1}  + \mathbf{b}_{z})
         \label{eq:u}
         \end{equation}
         \begin{equation}
         \widetilde{h_{t}} = tanh(\mathbf{W}_h\mathbf{x}_t + \mathbf{U}_h(\mathbf{r}_t \odot \mathbf{h}_{t-1}) + \mathbf{b}_{h})
         \end{equation}
          \begin{equation}
          r_{t} = \sigma(\mathbf{W}_r\mathbf{x}_t + \mathbf{U}_{r}\mathbf{h}_{t-1}  + \mathbf{b}_{r})
          \label{eq:r}
         \end{equation}  
         where $z_{t}$, $r_{t}$, $h_{t}$ and $\widetilde{h_{t}}$ are the signals over the update gate, reset gate, hidden state and candidate hidden state at time $t$, respectively; $\odot$ means element-wise product~\cite{chung2014empirical}.
      
\begin{figure}[t]
           \centering
           \includegraphics[width=\linewidth]{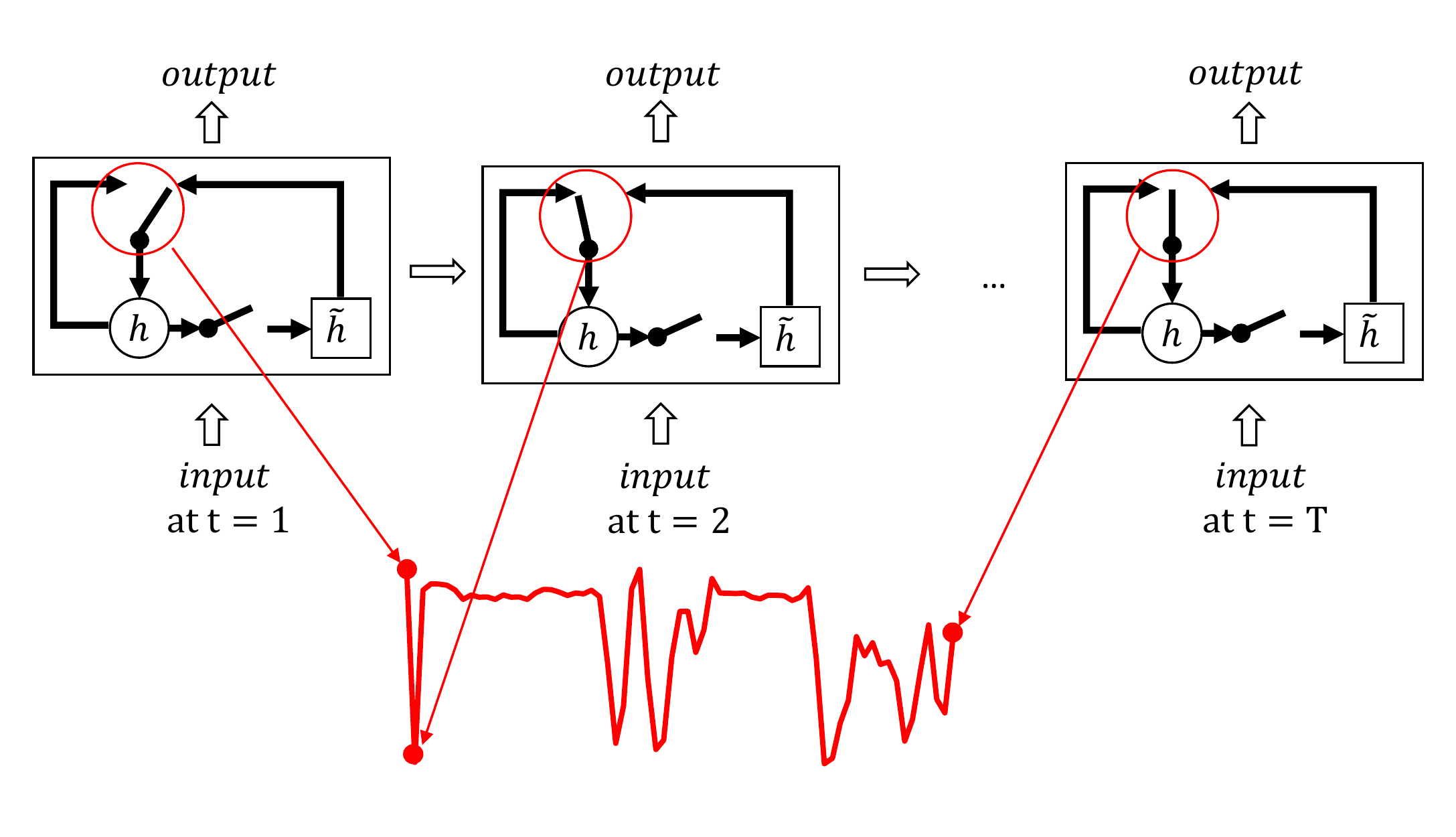}
           \caption{The schematic plot regarding how the gate activation signals may imply for the update gate of a GRU.}
           \label{fig:gas}
       \end{figure}

Here we wish to analyze the gate activations computed in equations (\ref{eq:f}), (\ref{eq:i}), (\ref{eq:o}), (\ref{eq:u}), (\ref{eq:r})~\cite{karpathy2015visualizing} and consider their temporal structures. 
For a GRNN layer consisting of $J$ gated units, we view the activations for a specific gate at time step $t$ as a vector $\mathbf{g}_t$ with dimensionality $J$, called \textit{gate activation signals} (GAS). 
Here $\mathbf{g}_t$ can be $\mathbf{h}_t$, $\mathbf{i}_t$, $\mathbf{o}_t$, $\mathbf{z}_t$ or $\mathbf{r}_t$ above.
Figure~\ref{fig:gas} is the schematic plot showing how GAS may imply for a gate in an gated unit.
        
     \subsection{Autoencoder GRNN}  
\begin{figure}[t]
           \centering
           \includegraphics[width=\linewidth]{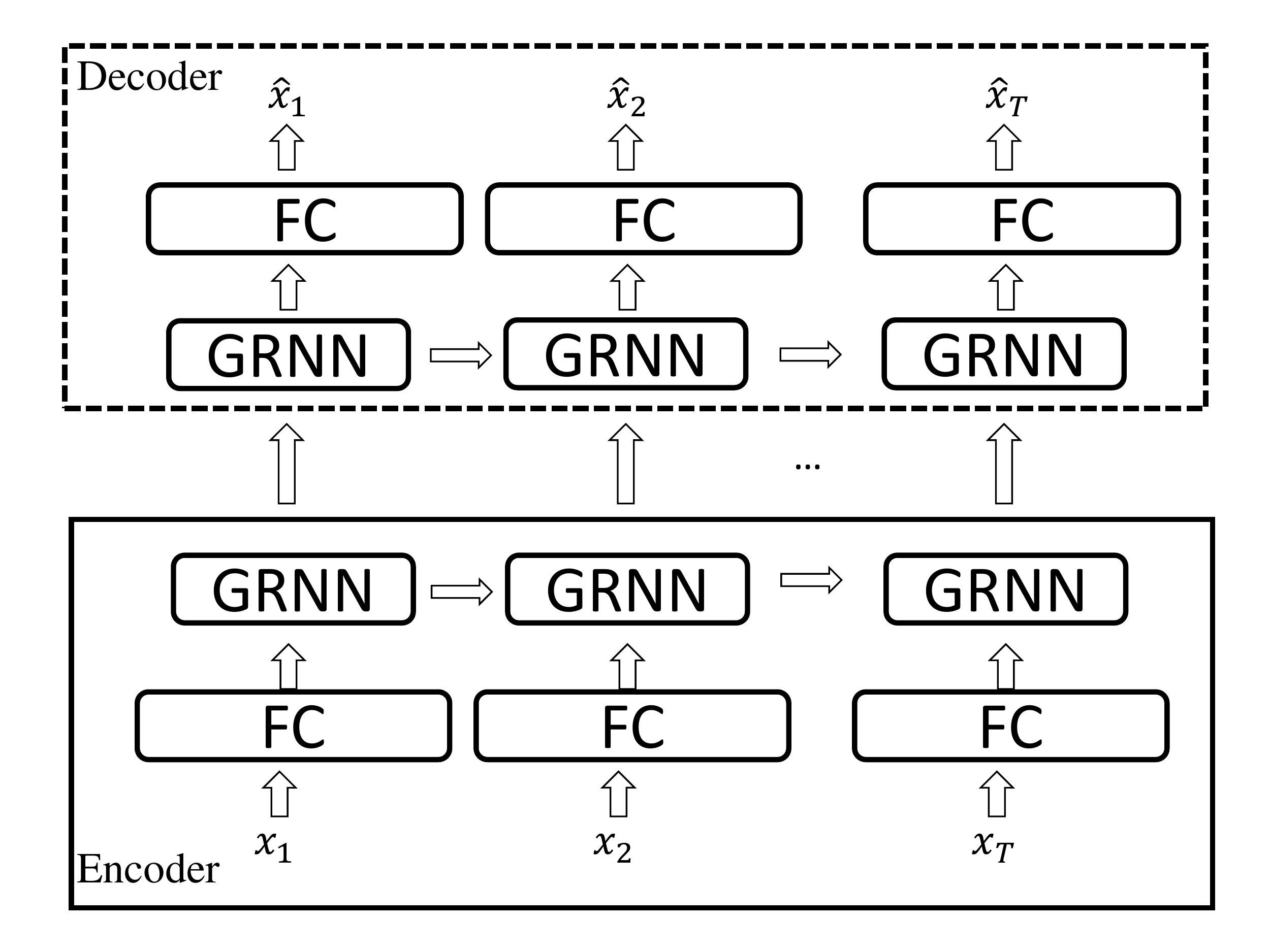}
           \caption{AE-GRNN structure consisting of an encoder and a decoder. The encoder consists of gated recurrent layers (GRNN) stacking on
           feed-forward fully-connected  layers (FC). Only one recurrent layer and one feed-forward layer are shown here for clarity. The structure of decoder is the mirrored structure of encoder.}
  \label{fig:AE_GRNN_fig}
       \end{figure}

Autoencoders can be trained in an unsupervised way, and have been shown to be very useful for many applications~\cite{deng2010binary}. We can have an autoencoder with GRNN as in Figure~\ref{fig:AE_GRNN_fig} called AE-GRNN here. Given an input utterance represented by its acoustic feature sequence $\{x_1,x_2,...,x_T\}$, at each time step $t$, AE-GRNN takes the input vector $x_t$, and produces the output $\hat{x}_t$, the reconstruction of $x_t$.
Due to the recurrent structure, to generate $\hat{x}_t$, AE-GRNN actually considers all information up to $x_t$ in the sequence, $x_1, x_2, ..., x_t$, or $\hat{x}_t = \textit{AE-GRNN}(x_{1}, x_{2}, ..., x_{t}$).
The loss function $\mathcal{L}$ of AE-GRNN in (\ref{eq:ae_loss_eq}) is the averaged squared $\ell$-2 norm for the reconstruction error of $x_t$.

        \begin{equation}
        \mathcal{L} = \sum_{n}^{N}\sum_{t}^{T_n}\frac{1}{d}\left \| {x}_{t}^{n} - \textit{AE-GRNN}(x_{1}^{n}, x_{2}^{n}, ..., x_{t}^{n}) \right \|^{2}
        \label{eq:ae_loss_eq}
        \end{equation}
where the superscript $n$ indicates the $n$-th training utterance with length $T_n$, and $N$ is the number of utterances used in training. $d$ indicates the number of dimensions of $x_t^n$.

\section{Initial Experiments and Analysis}
     \subsection{Experimental Setup}  
      We conducted our initial experiments on TIMIT, including 4620 training utterances and 1680 testing utterances. The ground truth phoneme boundaries provided in TIMIT are used here. We train models on the training utterances, and perform analysis on the testing ones.

      In the AE-GRNN tested, both the encoder and the decoder consisted of a recurrent layer and a feed-forward layer. The recurrent layers consisted of 32 recurrent units, while the feed-forward layers consisted of 64 neurons. We used ReLU as the activation function for the feed-forward layers~\cite{nair2010rectified}. The dropout rate was set to be 0.3~\cite{srivastava2014dropout}. The networks were trained with Adam~\cite{kingma2014adam}. The acoustic features used were the MFCCs of 39-dim with utterance-wise cepstral mean and variance normalization (CMVN) applied.
      
      \subsection{Initial Results and Observations}
\begin{figure}[t]
           \centering
           \includegraphics[width=1.1\linewidth]{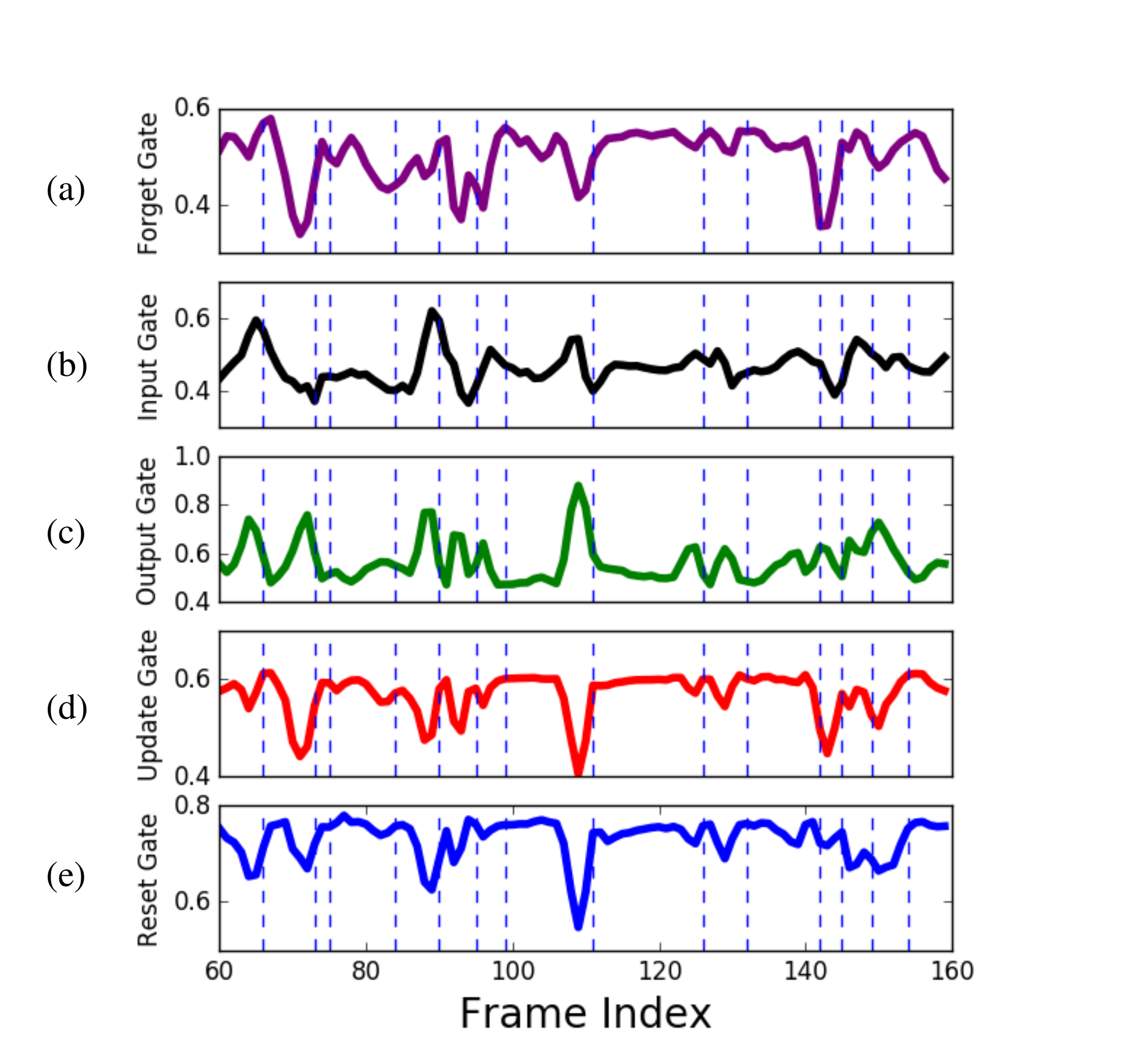}
           \caption{The means of the gate activation signals over all gated units for different gates with respect to the frame index.
           The blue dashed lines indicate the phoneme boundaries.}

           \label{fig:gas_frame}
      \end{figure}

      Figure~\ref{fig:gas_frame} shows the means of the gate activation signals over all gated units in the encoder of AE-GRNN with respect to the frame index, taken from an example utterance. The upper three subfigures (a)(b)(c) are for LSTM gates, while the lower two (d)(e) for GRU gates. 
      The temporal variations of GRU gates are similar to the forget gate of LSTM to some degree, and looks like the negative of the input gate of LSTM except for a level shift. More importantly, when compared with the phoneme boundaries shown as the blue dashed lines, a very strong correlation can be found. In other words, whenever the signal switches from a phoneme to the next across the phoneme boundary, the changes in signal characteristics are reflected in the gate activation signals. This is consistent to the previous finding that the sudden bursts of gate activations indicated that there were boundaries of phonemes in a speech synthesis task~\cite{wu2016investigating}.
       
       \subsection{Difference GAS}
\begin{figure}[t]
           \centering
           \includegraphics[width=\linewidth]{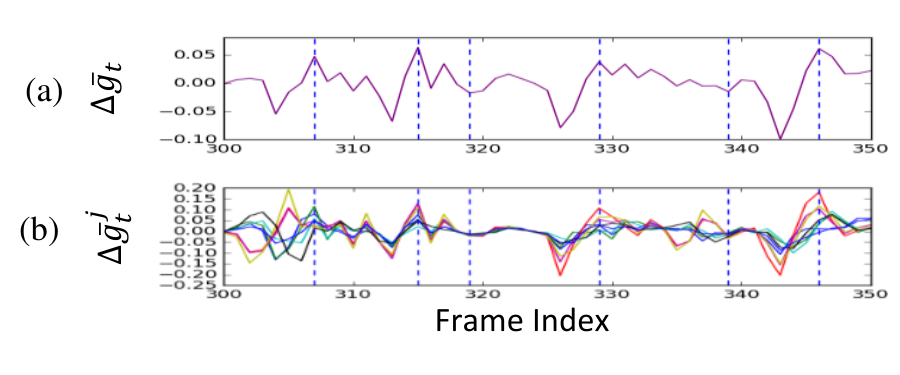}
           \caption{Subfigure (a) shows the plot of $\Delta\bar{g}_t$ for forget gate of LSTM for an example utterance over frame index. Subfigure (b) shows the plots of $\Delta{\bar{g}^j_t}$ with different colors (only shows $j=1$ to $8$ for clarity). The blue dashed lines indicate phone boundaries.}
            \label{fig:individual_gas}
       \end{figure}

       With the above observations, we define difference GAS as follows. 
       For a GAS vector at time index $t$, $\mathbf{g}_t$, we compute its mean over all units to get a real value $\bar{g}_t$. 
       We can then compute the difference GAS as the following:
       \begin{equation}
        \Delta\bar{g}_t = \bar{g}_{t+1} - \bar{g}_{t}
        \label{eq:delta_bar_g_eq}
        \end{equation}
        The difference GAS can also be evaluated for each individual gated unit for each dimension of the vector $\mathbf{g}_t$, 
        \begin{equation}
        \Delta\bar{g}_t^j = \bar{g}_{t+1}^j - \bar{g}_{t}^j
        \label{eq:delta_bar_g_eq_individual}
        \end{equation}
        where $g_t^j$ is the $j$-th component of the vector $\mathbf{g}_t$.
        We plotted the difference GAS and the individual difference GAS for the first 8 units in a GRNN over the frame index for an example utterance as in Figure~\ref{fig:individual_gas}.
        We see those differences bear even stronger correlation with phoneme boundaries shown by vertical dashed lines. All these results are consistent with the finding that the gate activations of forget gate over recurrent LSTM units in the same layer have close correlation with phoneme boundaries in speech synthesis~\cite{wu2016investigating}, although here the experiments were performed with AE-GRNN.
        
      \section{Phoneme Segmentation}
            
Because the GAS was found to be closely related to phoneme boundaries, we tried to use these signals to perform phoneme segmentation. The segmentation accuracy can be a good indicator to show the degree of the correlation between GAS and phoneme boundaries. In this section, we will first describe recurrent predictor model (RPM), an unsupervised phoneme segmentation approach,  which servers as the baseline. Then we will describe how to use GAS in phoneme segmentation.      
      
      \subsection{Baseline: Recurrent Predictor Model}
      RPM was proposed earlier to train GRNN without labeled data, and it was found the discontinuity on model outputs have to do with phoneme boundaries~\cite{drexler2016deep}\cite{DBLP:journals/corr/MichelRTD16}. An RPM has only the lower half of Figure~\ref{fig:AE_GRNN_fig}. The model output at time $t$, $\hat{x_{t}} = \textit{RPM}(x_1, x_2, ..., x_{t})$, is to predict the next input $x_{t+1}$. The loss function $\mathcal{L}$ used in training RPM is the averaged squared $\ell$-2 norm of prediction error,
       
        \begin{equation}
        \mathcal{L} = \sum_{n}^{N}\sum_{t}^{T_{n} -1}\frac{1}{d}\left \| x_{t+1}^{n} - \textit{RPM}(x_{1}^{n}, x_{2}^{n}, ..., x_{t}^{n}) \right \|^{2}
        \label{eq:rpm_loss_eq}
        \end{equation}
      which is actually parallel to (\ref{eq:ae_loss_eq}). The superscript $n$ indicates the $n$-th training utterance and $d$ indicates the number of dimensions of $x_t^n$.  Because frames which are difficult to predict or with significantly larger errors are likely to be phoneme boundaries, the error signals $E_t$ of RPM,
      \begin{equation}
        E_t = \frac{1}{d}\left \| x_{t+1}^{n} - \textit{RPM}(x_{1}^{n}, x_{2}^{n}, ..., x_{t}^{n}) \right \|^{2} 
        \label{eq:error_signal_eq}
        \end{equation}
        can be used to predict the phoneme boundary similar to GAS here. A time index is taken as a phoneme boundary if $E_t$ is a local maximum, that is $E_t >E_{t-1} $ and $E_t >E_{t+1}$, and $E_t$ exceeds a selected threshold. 
      
       \subsection{GAS for Phoneme Segmentation}
      
      From Figure~\ref{fig:individual_gas}, a direct approach to use GAS for phoneme segmentation is to take a time index as a phoneme boundary if $\Delta\bar{g}_t$ is a local maximum, that is $\Delta\bar{g}_t >\Delta\bar{g}_{t-1} $ and $\Delta\bar{g}_t >\Delta\bar{g}_{t+1}$, and $\Delta\bar{g}_t$ exceeds a selected threshold.
      
      GAS can also be integrated with RPM. Since RPM also includes GRNN within its structure, GAS can also be obtained and interpolated  with the error signals obtained in (\ref{eq:error_signal_eq}) as in (\ref{eq:interpolated_eq}), where $w$ is the weight. A time index is taken as a phoneme boundary if $I_t$ is a local maximum and exceeds a selected threshold.
      \begin{equation}
        I_t = (1-w)E_t + w\Delta\bar{g}_t
        \label{eq:interpolated_eq}
        \end{equation}

    \section{Experiments Results for Phoneme Segmentation}
    Here we take phoneme segmentation accuracy as an indicator to show the correlation between GAS and phoneme boundaries. The setup is the same as in Section 3.1. In the segmentation experiments, a 20-ms tolerance window is used for evaluation. 
    All GAS were obtained from the first recurrent layer. 
    Different segmentation results were obtained according to different thresholds, we report the best results in the following tables.
    
      \subsection{R-value Evaluation}
        It is well-known that the F1-score is not suitable for segmentation, because over segmentation may give very high recall leading to high F1-score, even with a relatively low precision\cite{DBLP:journals/corr/MichelRTD16}. In our preliminary experiments, a periodic predictor which predicted a boundary for every 40 ms gave F1-score 71.07 with precision 55.13 and recall 99.99, which didn't look reasonable. It has been shown that a better evaluation metric is the R-value~\cite{rasanen2009improved}, which properly penalized the over segmentation phenomenon. 
        The approach proposed in a previous work~\cite{rasanen2014basic} achieved an r-value 76.0, while the 40-ms periodic predictor only achieved 30.53. Therefore, we chose to use R-value on the performance measure.

      \subsection{Comparison among different gates}
      
       The R-values using different gates of LSTM and GRU are shown in Table~\ref{table:gates_exp}. The results for LSTM gates are consistent with the findings in the previous works~\cite{wu2016investigating}\cite{zaremba2015empirical}. 
        In LSTM, the forget gate clearly captures the temporal structure most related to phoneme boundaries.
        GRU outperformed LSTM which is also consistent with earlier results~\cite{chung2014empirical}\cite{zaremba2015empirical}. 
       The highest R-value is obtained with the update gate of GRU. The update gate in GRU is similar to the forget gate in LSTM. Both of them control whether the memory units should be overwritten. Interestingly, the reset gate of GRU achieved an R-value significantly higher than the corresponding input gate in LSTM. The reason is probably the location of reset gate. In GRU, the reset gate does not control the amount of the candidate hidden state independently, but shares some information of the update gate, thus has better access to more temporal information for phoneme segmentation~\cite{chung2014empirical}. 
       The update gate in GRU was used for extracting GAS in the following experiments.
       
    \begin{table}[th]
      \caption{The comparison between different gates in gated recurrent neural networks. }
      \label{table:gates_exp}
      \centering
      \begin{tabular}{lc}
        \toprule
        \textbf{Models}&\textbf{R-value}\\
        \midrule
          LSTM Forget Gate     & 79.15            \\
          LSTM Input Gate      & 70.75            \\
          LSTM Output Gate     & 61.97            \\
          GRU Update Gate      & \textbf{82.54}   \\
          GRU Reset Gate       &    78.94         \\
        \bottomrule
      \end{tabular}
    \end{table}

     \subsection{Comparison among different approaches}
     
\begin{figure}[t]
           \centering
           \includegraphics[width=\linewidth]{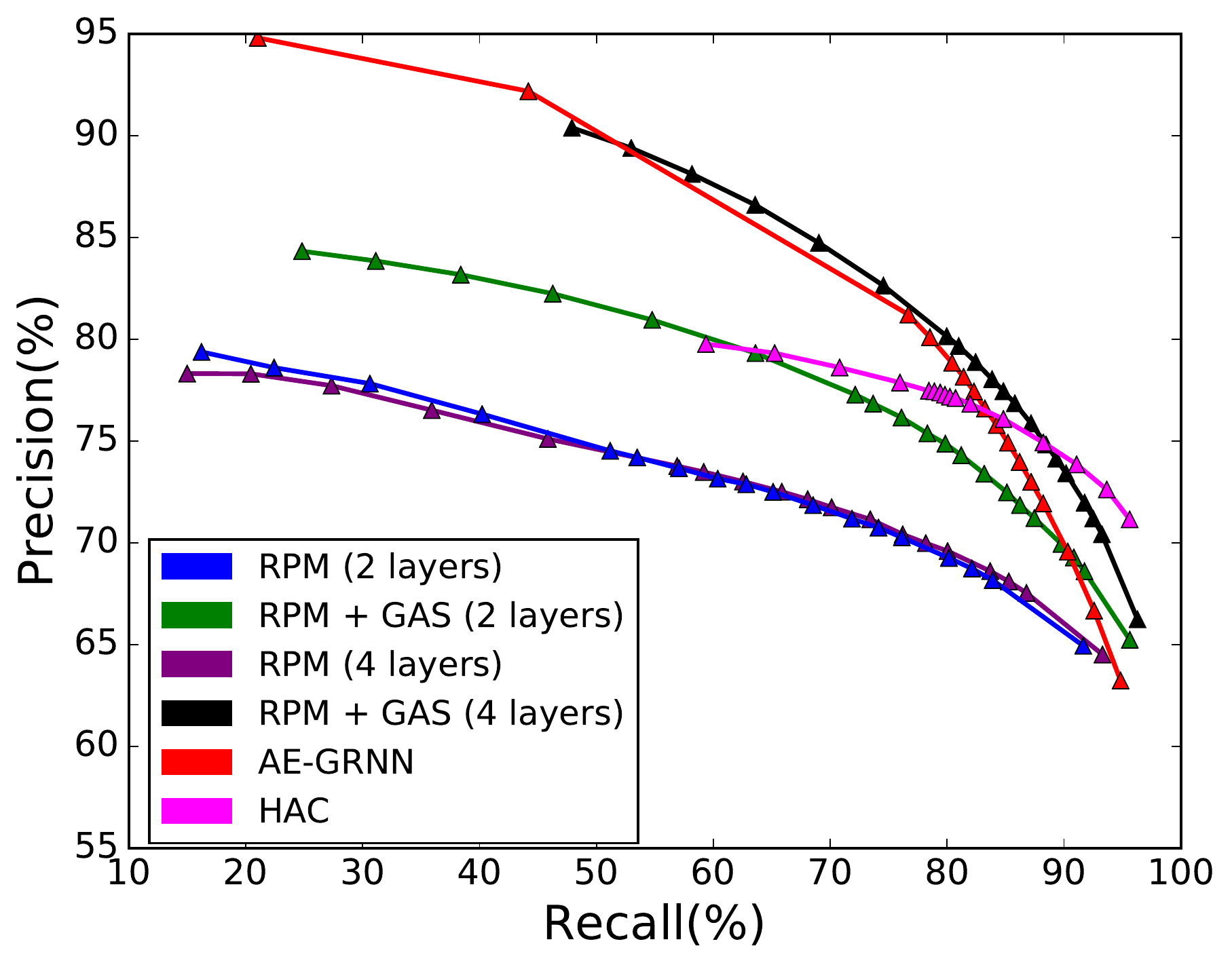}
           \caption{The Precision-Recall curves of different approaches in Table~\ref{table:main_exp}. Different markers on the curves stand for the results of different thresholds.}
           \label{fig:pr_curves}
    \end{figure}

          In Table~\ref{table:main_exp}, we compared the R-value obtained from the temporal information provided by RPM, without or with its GAS (rows (a)(b)(c)(d)).   
      The best result in Table~\ref{table:gates_exp} is in row(e), which was obtained with update gates of GRU in AE-GRNN.
          We considered two structures of RPM: the same as AE-GRNN (4 layers) or only use the encoder part (2 layers, a feed-forward layer plus a recurrent layer). The latter used the same number of parameters as AE-GRNN.
          We also tested the conventional approach of using hierarchical agglomerative clustering (HAC) as shown in row (f)~\cite{qiao2008unsupervised}\cite{chan2012unsupervised}. 
          We further added a white noise with 6dB SNR to the original TIMIT corpus (the right column).
          The last row (g) is for the periodic predictor predicted a boundary every 80 ms, serving as a naive baseline.
          Precision-Recall curves in Figure~\ref{fig:pr_curves} illustrate the overall performance on clean TIMIT corpus with different thresholds.
          
\begin{figure}[t]
           \centering
           \includegraphics[width=\linewidth]{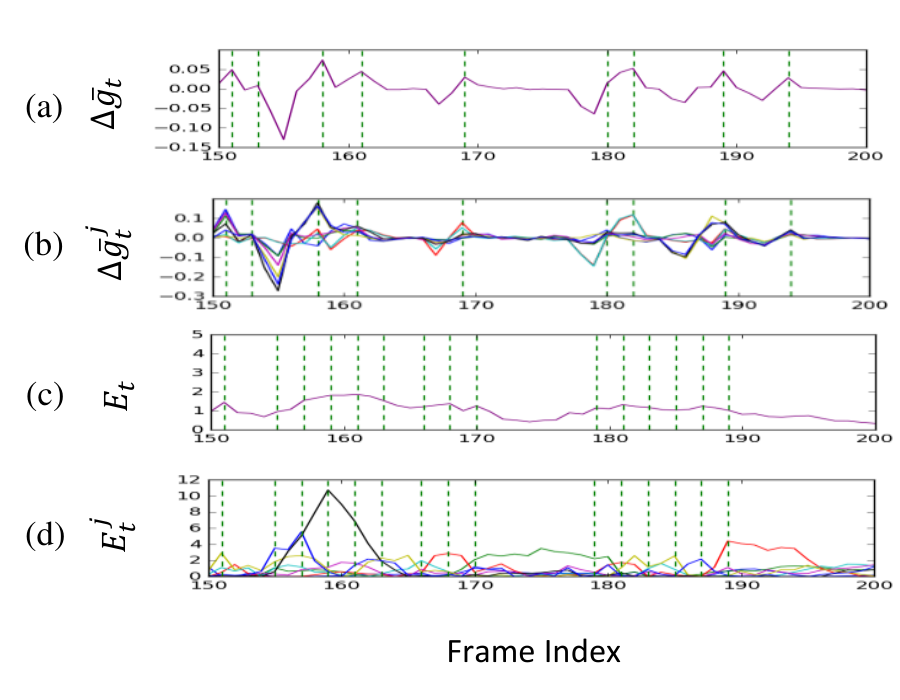}
            \caption{Subfigures (a) and (c) show the plots of $\Delta\bar{g}_t$ and $E_t$, respectively, for an example utterance over frame index. Subfigures (b) and (d) show the plots of $\Delta{\bar{g}^j_t}$ and $E^j_t$, respectively, with different colors (only show $j=1$ to $8$ for clarity). The green dashed lines indicate segmentation results.}
            \label{fig:rpm_detected_bounds}
    \end{figure}

           We found that the RPM performance was improved by the interpolation with GAS (RPM+GAS v.s. RPM). Larger improvements were gained when data became noisy.
    We further analyzed the segmentation results on clean corpus with the highest R-values of 2-layered RPM and AE-GRNN. We analyzed the results of AE-GRNN by observing $\Delta\bar{g}_t$ and $\Delta\bar{g}_t^j$. Likewise, we analyzed the results of RPM by observing $E_t$ and $E_t^j$, where $E_t^j$ indicates the squared prediction error in the $j^{th}$ dimension computed in (\ref{eq:error_signal_eq}). We showed their relations in Figure~\ref{fig:rpm_detected_bounds}. We see that the curve of $E_t$  is smooth and significantly different from the sharp curve of $\Delta\bar{g}_t$. The smoothness led RPM to suffer from over segmentation. The smoothness was caused by the fact that there were always a subset of $E_t^j$ which were significantly large. On the other hand, the curves of $\Delta\bar{g}_t^j$ are more consistent. The consistency enables curve of $\Delta\bar{g}_t$ to be sharp and thus AE-GRNN would not suffer from over segmentation. This explains why GAS are helpful here.     
    
      Also, we can see RPM alone didn't benefit much from adding more layers (rows (c) v.s. (a)).
     Providing if RPM is powerful enough to predict next frames better, there will be no error signals and thus no temporal information. This side effect balanced the advantage of increased model size. 
     Interestingly, not only GAS offered improvements (rows (b) v.s. (a) and rows (d) v.s. (c)), but with more layers, the interpolation with GAS achieved larger improvements (rows (d) v,s, (b)). 
    The best performances in both clean and noisy corpora are achieved by 4-layered RPM with GAS. 
    Last but not least, performance of approaches using GAS and HAC are more robust to noise.
          
    \begin{table}[th]
      \caption{The comparison of R-values for recurrent predictor model (RPM) without and with its internal GAS and GAS of AE-GRNN on clean and noisy TIMIT corpus.   
      The performance of hierarchical agglomerative clustering (HAC) and
      a periodic predictor are also included.}
      \label{table:main_exp}
      \centering
      \begin{tabular}{lccc}
        \toprule
        \textbf{Models}&\textbf{Clean}&\textbf{SNR-6dB}\\
        \midrule            
          (a) RPM  (2 layers)              & 76.02 &  73.7  \\
          (b) RPM + GAS (2 layers)      &    79.94      &  79.16 \\
          (c) RPM  (4 layers)             & 76.10  &  73.65  \\
           (d) RPM + GAS   (4 layers)      &  \textbf{83.16}  &  \textbf{81.54}\\
          (e) AE-GRNN    & 82.54  & 81.22\\
          (f) HAC &      81.61 & 80.41\\
          (g) Periodic Predictor  &  62.17  & 62.17\\
         \bottomrule
      \end{tabular}
    \end{table}
  
    \section{Conclusions}
     We show that the gate activation signals (GAS) obtained in an unsupervised fashion have temporal structures highly correlated with the phoneme changes in the signals, and this correlation was verified in the experiments for phoneme segmentation. 
     Also, our experiments also showed that GAS bring improvements to RPM without additional parameters.
     Like bottleneck features, GAS are obtained from the element of neural networks, instead of networks' outputs, and both of them are shown to bring improvements. With the promising results of GAS shown in the paper, we hope GAS can brought the same improvements as the ones brought by bottleneck features.
     
         
\newpage
\eightpt
\bibliographystyle{IEEEtran}

\end{document}